\documentclass{article}

\usepackage{arxiv}

\usepackage[utf8]{inputenc} % allow utf-8 input
\usepackage[T1]{fontenc}    % use 8-bit T1 fonts
\usepackage{hyperref}       % hyperlinks
\usepackage{url}            % simple URL typesetting
\usepackage{booktabs}       % professional-quality tables
\usepackage{amsfonts}       % blackboard math symbols
\usepackage{nicefrac}       % compact symbols for 1/2, etc.
\usepackage{microtype}      % microtypography
\usepackage{lipsum}
\usepackage{graphicx}
\usepackage{float}
\usepackage{tabularx}

\graphicspath{ {./images/} }

\usepackage[style=numeric,sorting=none,backend=biber]{biblatex}
\title{Dimensional Characterization and Pathway Modeling for Catastrophic AI Risks}

\author{
  Ze Shen Chin$^{1,2}$ \\
  $^1$Oxford Martin AI Governance Initiative \\
  $^2$AI Standards Lab \\
  \texttt{chinzeshen@gmail.com}
}

\date{August 2025}

\begin{document}

\maketitle

\begin{abstract}
Although discourse around the risks of Artificial Intelligence (AI) has grown, it often lacks a comprehensive, multidimensional framework, and concrete causal pathways mapping hazard to harm. This paper aims to bridge this gap by examining six commonly discussed AI catastrophic risks: CBRN, cyber offense, sudden loss of control, gradual loss of control, environmental risk, and geopolitical risk. First, we characterize these risks across seven key dimensions, namely intent, competency, entity, polarity, linearity, reach, and order. Next, we conduct risk pathway modeling by mapping step-by-step progressions from the initial hazard to the resulting harms. The dimensional approach supports systematic risk identification and generalizable mitigation strategies, while risk pathway models help identify scenario-specific interventions. Together, these methods offer a more structured and actionable foundation for managing catastrophic AI risks across the value chain.
\end{abstract}

\section{Introduction}
The rapidly developing field of artificial intelligence (AI) has raised the discourse on risks ranging from immediate and tangible harms to catastrophic threats \cite{Bengio_2024}. While there have been several attempts to create taxonomies of risks, there has yet to be a way to contextualize or characterize AI risks holistically.

Separately, while the discourse on AI risks includes widespread concern among prominent figures about the risk of human extinction from AI \cite{caisairisk2023}, such risks are mostly articulated in comparatively vague and \textit{a priori} terms \cite{dubber2025militaryaicyberagents}. There is a notable lack of concrete, step-by-step processes detailing how AI systems could transition from their current capabilities to a state where they pose a catastrophic threat to humanity, with clear causal pathways. 

This paper is motivated by a fundamental need to bridge these conceptual gaps. We aim to:
\begin{itemize}
    \item Provide a framework for risk characterization according to risk dimensions.
    \item Enable a deeper understanding of risks by mapping out scenarios with more granular causal pathways.
    \item Allow for the design and implementation of assessment and mitigation measures, both for broad categories of risks, as well as specific scenarios at each stage of these identified pathways.
\end{itemize}

This work also responds to the recognized need for more detailed threat models in AI risk assessment, as highlighted in \textit{Open problems in technical AI governance} \cite{reuel2025openproblemstechnicalai}. Although existing taxonomies already provide various ways of categorizing risks, our contribution consolidates the different types of risk into several dimensions and expands on the pathways of broad classes of catastrophic AI risks that are commonly discussed.

\section{Related Work}
\subsection{Overview of AI existential or catastrophic risks}
There is a lot of existing work that gives overviews of broad pathways of AI existential or catastrophic risks. \cite{dsit2024future} outlines three broad pathways: misalignment, single point of failure, and overreliance. \cite{dsit2025frontier} also discusses frontier AI risks under three headings: societal harms, misuse, and loss of control. In addition, \cite{dsit2025safety} discusses advanced AI risks under risks from malicious use, risks from malfunctions, and systemic risks. \cite{eisenpress2024catastrophic} similarly discusses three types of catastrophic risks from AI: risks from bad actors, systemic risks, and rogue AI. \cite{hendrycks2023overviewcatastrophicairisks} organizes catastrophic AI risks into four categories: malicious use, AI race, organizational risks, and rogue AIs. \cite{Chan_2023} explores more specifically harms from increasingly agentic algorithmic systems which include systemic and delayed harms, collective disempowerment, and harms yet to be identified. \cite{shah2025approachtechnicalagisafety} considers four risk areas: misuse, misalignment, mistakes, and structural risks.

In the literature that covers a broad overview of existential or catastrophic risks, the coverage and categorization of these risks seem to be fairly subjective. \cite{kasirzadeh2025typesaiexistentialrisk} describes two types of AI existential risk: decisive and accumulative. \cite{swoboda2025examiningpopularargumentsai} unpacks existential AI risk scenarios into two pathways: accidents and/or misuse of AI by malicious actors, and AI pursuing goals misaligned with human values. \cite{critch2023tasrataxonomyanalysissocietalscale} explores a taxonomy based on accountability, proposing six types of risks: diffusion of responsibility, “bigger than expected”, “worse than expected”, willful indifference, criminal weaponization, and state weaponization. \cite{brundage2024malicioususeartificialintelligence} explores malicious use of AI structured around three security domains: digital security, physical security, and political security. \cite{Bucknall_2022} explores existential risk factors including general risk factors, nuclear weaponry, pandemics and biotechnology, climate change, natural risks, and unaligned AGI. 

There are also attempts to systematically review existential or catastrophic risks and organize them according to certain dimensions. \cite{yampolskiy2015taxonomypathwaysdangerousai} classifies pathways to dangerous AI according to two dimensions: timing (pre-deployment or post-deployment) and causes (external or internal causes). \cite{turchin2020catastrophicAIRisks} classifies global catastrophic risks connected to AI into two different dimensions: AI level (including narrow AI, young AI, or mature AI) and agency (including human agency, AI’s agency, relationship of two agents: AI and human, many agents, or no agency). 

Beyond the scope of catastrophic risks, several other works attempt to taxonomize risks or harms under various different bases. This includes \cite{slattery2025airiskrepositorycomprehensive} and \cite{uuk2024taxonomysystemicrisksgeneralpurpose} based on systematic review of academic literature, \cite{10.1145/3531146.3533088} based on both horizon-scanning workshops and discussions as well as a literature review, \cite{abercrombie2024collaborativehumancentredtaxonomyai} based on other taxonomies, topic experts, and reported incidents or issues, \cite{oecd2023aiincident} based on definitions included in key standards and legislation, \cite{hoffmann2023aiharm} based on existing incidents, and \cite{zeng2024airiskcategorizationdecoded} based on government policies. 

\subsection{Implications on AI risk management}
Discourse around catastrophic or existential risks sometimes lead to different recommendations. \cite{Barrett_2016} presents a model of major pathways to artificial superintelligence (ASI) catastrophe, and recommends that the entire study of ASI catastrophe risk to be a significant research priority. On the other hand, \cite{sotos2021unimportancesuperintelligence} argues that resources should be preferentially applied to mitigating the risk of peripheral systems and savant software, instead of hypothetical risks of superintelligence. 

Certain AI standards or policies have included certain risks for the purpose of risk management, which differ according to geographical regions. For example, in the United States, \cite{nist2023airmf1} discusses AI risks but does not prescribe specific risk types, where it only describes a list of characteristics of trustworthy AI systems; while \cite{nist2021taxonomyairisk} discusses a taxonomy of risk with three categories: technical design attributes, how AI systems are perceived, and guiding policies and principles. In China, \cite{tc260aisafetygovernance2024} classifies safety risks into inherent safety risks (including risks from models and algorithms, risks from data, and risks from AI systems) and safety risks in AI applications (including cyberspace risks, real-world risks, cognitive risks, and ethical risks). In the European Union (EU), \cite{ec2025gpai_codecpractice3} lists four types of systemic risks for the General-purpose AI (GPAI) Code of Practice, namely cyber offense, chemical, biological, radiological and nuclear (CBRN) risks, loss of control, cyber offense, and harmful manipulation.

With the understanding that certain capabilities are precursors to risks, management of these risks have also been translated into conducting capabilities assessment in frontier AI labs. Safety frameworks from frontier AI companies such as \cite{anthropicasl32025}, \cite{openaipreparedness2025}, \cite{deepmindfrontiersafety2025}, \cite{metafrontierai2025}, and \cite{microsoftfrontiergovernance2025} generally include assessment of certain capabilities as part of the risk management process. The Frontier Model Forum has thus included CBRN threats, advanced cyber threats, and advanced autonomous behavior threats within the current domain of consensus \cite{fmfrisktaxonomy2025}. Nevertheless, the risk management landscape downstream of frontier models is relatively less mature.

\section{Definitions}
The following definitions are used throughout this paper.

\textbf{Risk}: effect of uncertainty on objective, usually expressed in terms of risk sources, potential events, their consequences, and their likelihood (as specified in \cite{iso310002018}).

\textbf{Hazard}: a potential source of harm (as specified in \cite{iso149712019})

\textbf{Risk source}: element which alone or in combination has the potential to give rise to risk (as specified in \cite{iso310002018}).

\textbf{Event}: occurrence or change of a particular set of circumstances (as specified in \cite{iso310002018}).

\textbf{Consequence}: outcome of an event affecting objectives (as specified in \cite{iso310002018}).

\textbf{Catastrophic risk}: the risk of widespread and significant harm, such as several million fatalities or severe disruption to the social and political global order \cite{koessler2023riskassessmentagicompanies}.

\textbf{Existential risk}: a risk that threatens the destruction of humanity’s long-term potential \cite{ordprecipice2021}. This is a subset of catastrophic risk.

\section{Methodology}
We explore and select broad types of catastrophic risks that are commonly discussed in the literature. They include:
\begin{itemize}
    \item CBRN
    \item Cyber offense
    \item Sudden loss of control
    \item Gradual loss of control
    \item Environmental risk
    \item Geopolitical risk
\end{itemize}

The first three risks (CBRN, cyber offense, sudden loss of control) are selected as they represent the basis for the capability assessments that are typically conducted by frontier labs \cite{fmfrisktaxonomy2025}. The remaining three were included as they represent distinct combinations of risk dimensions and are increasingly regarded as credible catastrophic threats.

For each risk, we characterize the risk in terms of its risk dimensions and conduct simple risk pathway modeling by mapping out causal pathways to harm. These are distinct but complementary analyses: risk dimensions provide a way to characterize the risk according to different attributes, and identify broadly applicable risk management measures relevant to those attributes; while risk pathway modeling provides a way to visualize how risks can concretely unfold into a harm, and identify specific risk management measures according to those pathways. We then describe an example of a historical precedent that is analogous to the risk, and discuss its similarities and differences.

\subsection{Risk dimensional characterization}

The dynamic nature of AI development compels a shift from static risk categories to a more nuanced dimensional approach \cite{engin2025adaptivecategoriesdimensionalgovernance}. This integrated dimensional and categorical thinking finds precedent in fields like psychiatric diagnosis \cite{widigerdimensions2005}, \cite{CHMURAKRAEMER200417}, developmental psychology \cite{sieglervariability2007}, and cognitive science \cite{fazekasmultifactor2018}, where it emerged to address complexity, non-linear patterns, and context dependency that traditional categorical models could not sufficiently capture. Applying this to AI risks allows a better representation of the multi-faceted causes and factors that lead to these risks.

Some of these dimensions relevant to AI risks have been discussed in the literature. For example, \cite{oecd2023aiincident} discusses several dimensions of harm, namely type, level of severity, scope (type of harmed entity), geographic scale, tangibility, quantifiability, materialisation, reversibility, recurrence, impact, and timeframe. \cite{ec2025gpai_codecpractice3} considers several nature of systemic risks to be used to inform the selection of risks, including specific to advanced capabilities, significant impact, high velocity, compounding or cascading, difficult or impossible to reverse, and asymmetric impact. \cite{gipiškis2024risksourcesriskmanagement} proposes a taxonomy of harm dimensions, including direct harm domains and negative externality domains; as well as a taxonomy of risk source dimensions, including intent, entity, failure dynamics, technical attributes, and stage of risk emergence. 

In the following sections, we adopt and expand on these dimensions from the literature, and then discuss the relevant risk management measures for the various attributes under each dimension. 

\subsection{Risk pathway modeling}
Risk modeling is a common technique in risk assessment, however, it has no universally accepted definition. In the finance industry, \cite{deloitteriskmodeling} defines a risk model as “a mathematical representation of a system, commonly incorporating probability distributions”. \cite{experianriskmodeling2025} defines it as “a representation of a particular situation that’s created specifically for the purpose of assessing risk”, which is then used to “evaluate the potential impacts of different decisions, paths and events”.

\cite{nistsp80030risk2012} defines it as “key component of a risk assessment methodology that defines key terms and assessable risk factors”, where risk factors include threat, vulnerability, impact, likelihood, and predisposing condition.

In the context of AI risk management, there is less emphasis on risk modeling being quantitative. \cite{campos2025frontierairiskmanagement} defines it as “the systematic process of analyzing how identified risks could materialize into concrete harms”, which “involves creating detailed step-by-step scenarios of risk pathways that can be used to estimate probabilities and inform mitigation strategies”. \cite{ec2025gpai_codecpractice3} defines it similarly, as “a structured process aimed at specifying pathways through which a systemic risk stemming from a model might materialise”. They also note that it is “often used interchangeably with the term 'threat modeling'”, but ‘risk modeling’ is used instead because “the term 'threat modeling' has a specific meaning in cybersecurity”. We believe the distinction is important, as ‘threat modeling’ implies the presence of a threat actor, but there are risks related to AI that do not necessarily involve a clear threat actor. 

Nevertheless, in the following section, we will use the term “risk pathway modeling” as introduced by \cite{wisakanto2025adaptingprobabilisticriskassessment}, defined as “modeling of the step-by-step progressions of risk from a system’s source aspects to terminal aspects”, to reflect the focus on the mapping of pathways instead of the creation of a model that aims to be representative of the risk in a quantitative way. 

We first model each risk by listing the associated hazard, event, and consequence, along with other relevant risk sources for an example scenario. While there is some subjectivity in distinguishing hazards from other risk sources - as a hazard is a potential cause of harm whose release leads to an event and subsequent consequences - we have designated model capabilities as hazards due to their dual-use nature, where applicable. This aligns with the focus on capability assessments by frontier model developers, which often reflects hazard analysis in safety engineering \cite{fmfcapability2025}.

Next, based on an example scenario illustrative of each risk, we map out the causal pathways in the form of a flowchart, where we include the relevant risk sources and how they lead to concrete harms. The harms considered broadly include areas such as physical harms, emotional harms, and financial or economic harms, environmental harms, and loss of autonomy. 

\section{Risk dimensions}
In this section, we discuss several dimensions of risks and list out the risk management measures broadly relevant to the specific attributes of those dimensions. The risk management measures listed may be of different phases of the risk management process, such as risk identification, risk analysis, risk evaluation, risk mitigation, and monitoring and review.

\subsection{Intent}
Within the context of intent, AI risks are often categorized as either an accident or an act of misuse \cite{delaney2024mappingtechnicalsafetyresearch}.

\begin{table}[H]
\centering
\caption{Types of intent with the relevant risk management measures}
\begin{tabularx}{\textwidth}{|p{5cm}|X|}
\hline
\textbf{Types of intent} & \textbf{Risk management measures} \\
\hline
Intentional &
$\bullet$ Know your customer (KYC) practices \cite{UK_Gov2023EmergingProcesses} \newline
$\bullet$ Detect and disrupt malicious uses \cite{OpenAI2024StateThreatActors} \newline
$\bullet$ Strengthen model security \cite{grosse2024practicalthreatmodelsartificial, yazmyradov2024aisecurity, harris2025aisupplychain} \newline
$\bullet$ Strengthen cybersecurity \cite{ee2024adaptingcybersecurityframeworksmanage, degregorio2025mitigatingcyberriskage} \newline
$\bullet$ Build safety cases for safeguards against misuse \cite{clymer2025examplesafetycasesafeguards} \newline
$\bullet$ Govern as dual-use technology \cite{wasil2024governingdualusetechnologiescase} \\
\hline
Unintentional &
$\bullet$ Improve accountability mechanisms \cite{richards2025incidentsinsightspatternsresponsibility} \newline
$\bullet$ Failure cause analysis for AI incidents \cite{pittaras2022taxonomicfailurecauseanalysis} \\
\hline
\end{tabularx}
\label{tab:intent}
\end{table}

\subsection{Competency}
Risks can arise due to model capabilities or model failures \cite{yampolskiy2016artificialintelligencesafetycybersecurity}. This can also be framed as competence-based or incompetence-based hazards \cite{wisakanto2025adaptingprobabilisticriskassessment}, where models either succeed at something we do not want or they fail at something we want. 

\begin{table}[H]
\centering
\caption{Types of competency with the relevant risk management measures}
\begin{tabularx}{\textwidth}{|p{5cm}|X|}
\hline
\textbf{Types of competency} & \textbf{Risk management measures} \\
\hline
Competent (capability) &
$\bullet$ Capability evaluations \cite{shevlane2023modelevaluationextremerisks} \\
\hline
Incompetent (failure) &
$\bullet$ Create trustworthy AI \cite{li2022trustworthyaiprinciplespractices} \\
\hline
\end{tabularx}
\label{tab:competency}
\end{table}

\subsection{Entity}
Risks can emerge from humans or from AIs, but there is a wide spectrum in between the two, where an AI can operate with different levels of autonomy \cite{VAGIA2016190}, \cite{844354}.

\begin{table}[H]
\centering
\caption{Types of entity with the relevant risk management measures}
\begin{tabularx}{\textwidth}{|p{5cm}|X|}
\hline
\textbf{Types of entity} & \textbf{Risk management measures} \\
\hline
Humans &
$\bullet$ Capability evaluations \cite{attardfrost2024ethicsaivaluechains} \\
\hline
AI &
$\bullet$ Development of non-agentic AIs \cite{bengio2025superintelligentagentsposecatastrophic} \newline
$\bullet$ Visibility into AI agents \cite{chan2024visibilityaiagents} \newline
$\bullet$ Governance frameworks for agentic AI systems \cite{openaiagenticpractices2023}\\
\hline
Combination of humans and AI &
$\bullet$ Frameworks for human-AI collaboration \cite{mohsin2025unifiedframeworkhumanai} \\
\hline
\end{tabularx}
\label{tab:entity}
\end{table}

\subsection{Polarity}

AI risks are commonly viewed from the perspective of one independent AI tool or agent, where the misuse, malfunction, or misalignment of an AI leads to harm. However, there have also been framing of risks arising from multi-agent interactions, where risks arise from the interactions between multiple parties in which the cause of harm cannot be traced back to a single agent \cite{drexlercais2019}, \cite{CAIF_1}.

\begin{table}[H]
\centering
\caption{Types of polarity with the relevant risk management measures}
\begin{tabularx}{\textwidth}{|p{5cm}|X|}
\hline
\textbf{Types of polarity} & \textbf{Risk management measures} \\
\hline
Single agent &
(as per other risk management measures) \\
\hline
Multi-agent &
$\bullet$ Map interactions between agents, similar to ecosystem graphs \cite{bommasani2023ecosystemgraphssocialfootprint} \newline
$\bullet$ Develop infrastructure for AI agents \cite{chan2025infrastructureaiagents} \newline
$\bullet$ Threat modeling for multi-agent systems \cite{owaspmasthreats2025}
\\
\hline
\end{tabularx}
\label{tab:polarity}
\end{table}

\subsection{Linearity}
The realization of risks can follow either linear or non-linear causal pathways. Drawing from Perrow's Normal Accident Theory, even if individual components appear to operate as intended, highly complex and tightly coupled systems are inherently susceptible to "normal accidents" where unforeseen interactions of minor failures lead to catastrophic outcomes \cite{69444197-437a-3875-b762-d56c857caebc}. Non-linear causal pathways are particularly likely in complex systems that contain unpredictable scaling and emergence, feedback loops, cascading effects, and tail risks \cite{kolt2025lessonscomplexitytheoryai}. Risks that arise primarily from non-linear pathways are also often described as structural risks \cite{zwetslootstructure2019}, \cite{kilian2025accidentsmisusedecodingstructural}. For example, in a highly optimized global supply chain, a minor disruption could trigger unforeseen feedback loops, with the potential of causing rapid cascading failures and a disproportionate system-wide collapse.

\begin{table}[H]
\centering
\caption{Types of linearity with the relevant risk management measures}
\begin{tabularx}{\textwidth}{|p{5cm}|X|}
\hline
\textbf{Types of linearity} & \textbf{Risk management measures} \\
\hline
Linear &
$\bullet$ Use of traditional risk management techniques \cite{barrett2025airiskmanagementstandardsprofile} \\
\hline
Non-linear (systemic or structural) &
$\bullet$ Use of risk management techniques like Systems-Theoretic Process Analysis (STPA) \cite{mylius2025systematichazardanalysisfrontier} \newline
$\bullet$ Use of structural approach to inform policies \cite{vanderloeff2019aiethicssystemicissues} \newline
$\bullet$ Regulatory focus on structural constraints and dependencies that prevent harm \cite{wang2025distinguishingpredictivegenerativeai}
\\
\hline
\end{tabularx}
\label{tab:linearity}
\end{table}

\subsection{Reach}
Some risks exhibit an internalized reach, meaning their consequences are predominantly confined to individuals or entities directly involved in the chain of activity. In contrast, risks can have an external reach (or spillover reach), where their effects extend beyond those directly involved. These broader impacts often manifest as externalities, defined in economic theory as costs or benefits imposed on a third party who is not directly engaged in the activity or transaction causing the risk \cite{mankiwprinciples2020}.

\begin{table}[H]
\centering
\caption{Types of reach with the relevant risk management measures}
\begin{tabularx}{\textwidth}{|p{5cm}|X|}
\hline
\textbf{Types of reach} & \textbf{Risk management measures} \\
\hline
Internalized &
(as per other risk management measures) \\
\hline
Externalized (spillover) &
$\bullet$ Algorithmic impact assessment \cite{oecdaiasstool2025} \newline
$\bullet$ Systems thinking to account for externalities \cite{nokhiz2025rethinkingoptimizationsystemsbasedapproach} \newline
$\bullet$ Mechanism design \cite{zheng2025mechanismdesignauctionsexternalities}
\\
\hline
\end{tabularx}
\label{tab:reach}
\end{table}

\subsection{Order}
First-order effects of AI risks are more commonly discussed in the literature, whereas second-order effects are often neglected \cite{tan2022risksmachinelearningsystems}. Second-order risks arise as unintended consequences to other initial first-order effects. For example, if a first-order effect is mass unemployment due to AI automation, a second-order effect might be the political instability and societal breakdown that arises in response to mass unemployment.

Second-order risks are distinct from both externalities and non-linear risks: externalities can be first-order effects that impact a third party; while non-linear risks can also be first-order effects that involve complex systems with feedback loops.

\begin{table}[H]
\centering
\caption{Types of order with the relevant risk management measures}
\begin{tabularx}{\textwidth}{|p{5cm}|X|}
\hline
\textbf{Types of order} & \textbf{Risk management measures} \\
\hline
First-order &
(as per other risk management measures) \\
\hline
Second-order (or more) &
$\bullet$ Real world evaluation ecosystem \cite{schwartz2025realitychecknewevaluation}
\\
\hline
\end{tabularx}
\label{tab:order}
\end{table}

\section{Risks}
In this section, for each of the AI-related catastrophic risks, we discuss its risk dimensions and sketch out a risk pathway model. In some cases, certain dimensions are more central to the risks, while other dimensions may not be specifically relevant to those risks. Where the latter applies, we denote “variable” under those specific dimensions.

\subsection{CBRN}
Chemical, biological, radiological, and nuclear (CBRN) risks are broad classes of threats that have the potential to cause harm to a large number of people. Explosives are also sometimes included in this category, often referred to as CBRNE.

\subsubsection{Risk dimensions}
\begin{itemize}
    \item \textbf{Intent}: Intentional
    \item \textbf{Competency}: Competent
    \item \textbf{Entity}: Humans
    \item \textbf{Polarity}: Single-agent
    \item \textbf{Linearity}: Linear
    \item \textbf{Reach}: Internalized
    \item \textbf{Order}: First-order
\end{itemize}

The key characteristic of CBRN risk is that it stems from misuse of capable models with a direct pathway to harm, where a malicious actor is able to carry out consequential attacks more efficiently and effectively with the help of AI. 

\subsubsection{Risk pathway model}
\begin{itemize}
    \item \textbf{Hazard}: Model with CBRN capabilities
    \item \textbf{Event}: A CBRN agent is released
    \item \textbf{Consequence}: Mass injury and loss of lives
\end{itemize}

Here, we designate a model CBRN capabilities as the hazard, as these capabilities are dual-use which may or may not lead to harm. For example, a model with biological capabilities can both be used for conducting beneficial medical research. However, when in the hands of a malicious actor, such models can be used to aid development of CBRN weapons, which can lead to the event of the release of a CBRN agent that has potentially catastrophic consequences.

Below, we explore a subset of CBRN risk: the risk of the creation and release of biological weapons. As there has yet to be credible claims that AI models are suggesting previously-unknown pathways to bioweapons development \cite{berkepathogen2023}, the role of AI here is primarily in terms of uplift, where certain steps within the existing pathway to harm may be enhanced or made more accessible by AIs. These AIs currently fall into two broad categories: large language models (LLMs) and biological design tools (BDTs) \cite{sandbrink2023artificialintelligencebiologicalmisuse}.

The extent to which LLMs alone contribute to bioweapons uplift is debated. According to \cite{peppin2025realityaibiorisk}, the available literature at the time did not support the notion that the use of publicly available LLMs can significantly increase biorisk; while \cite{brent2025contemporaryaifoundationmodels} finds that recent LLMs can provide significant guidance to motivated actors in developing bioweapons. 

More broadly, the combination of LLMs and biological tools (BTs) presents clearer risks. Building on a risk chain developed by \cite{sandbergriskchain2020}, \cite{nelsonbioweapon2023} finds that capabilities from both LLMs and BTs have the potential to enable capabilities at multiple steps in the risk chain.

\begin{figure}[htbp]
    \centering
    \includegraphics[width=0.8\textwidth]{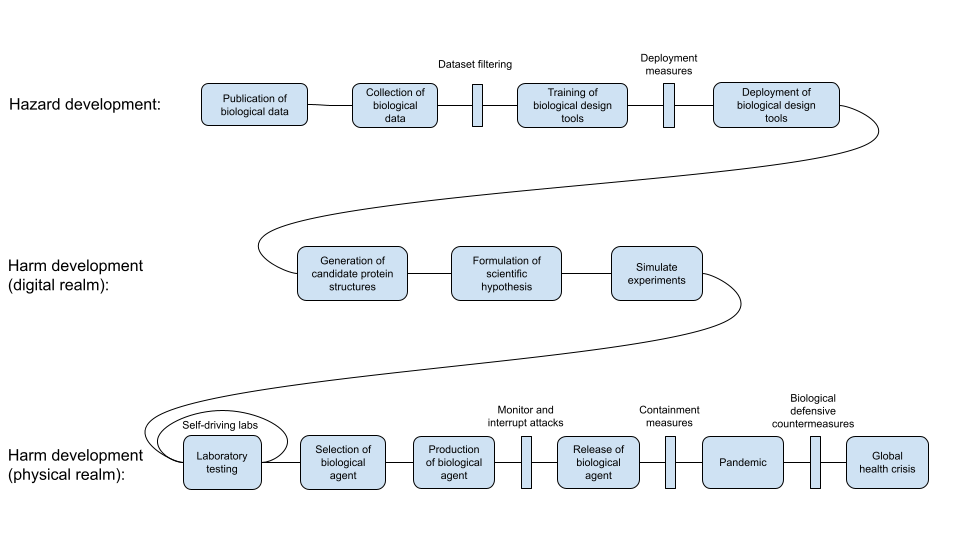}
    \caption{Risk pathway model for a biological risk scenario}
    \label{fig:biological_risk}
\end{figure}

We explore a pathway of biological risks based on \cite{randaiweapons2025} and \cite{brent2025contemporaryaifoundationmodels}, as per Figure \ref{fig:biological_risk}, where the hazard, a biological design tool, is developed and deployed. A malicious actor can then use it to help design, produce, and release a biological agent, thereby causing harm. Risk management measures can be placed upstream to reduce the capabilities of the BDT, and they can be placed downstream to reduce the likelihood or severity of a pandemic. 

\subsubsection{Analogous historical precedence}
The 2001 US anthrax attack is one of the worst biological attacks in history, where five people were killed and 17 others infected, with several senators being victims of the attack. Anthrax, an infection caused by the bacterium Bacillus anthracis, is deadliest when spread through inhalation of anthrax spores \cite{CDC2025Anthrax}. Investigations conclude that the perpetrator, who had access to highly sophisticated lab equipment, possessed the knowledge and ability of growing, harvesting, storing, and drying highly purified spores used in the mailings \cite{DOJ2010Amerithrax}. 

While modern AI was not involved in the 2001 attacks, it is believed that AI-driven biotech will make bioweapons easier and cheaper to develop over the next decade \cite{wefglobalrisks2025}. This raises concerns that similar attacks could be carried out by individuals with less specialized knowledge and resources.

\subsubsection{Other similar risks}
There are other types of risks that share similar risk dimensions but arise from very different pathways. For example, the development and deployment of nanoweapons which may lead to catastrophic harms \cite{bostromextinction2002}. Separately, the use of AI-enabled surveillance and control employed by state or non-state actors could facilitate authoritarian regimes and the eventual loss of autonomy \cite{bullock2025agigovernmentsfreesocieties}, \cite{aigiauthoritarianism2025}. 

\subsection{Cyber offense}
Cyber risks, especially in the context of cyber offense, are an existing threat that may be exacerbated by AI. \cite{zhu2025teamsllmagentsexploit} demonstrated that teams of LLM agents can exploit zero-day vulnerabilities when given a description of the vulnerability and toy capture-the-flag problems. While cyber risks are not typically regarded as catastrophic, \cite{dubber2025militaryaicyberagents} argues that cyberwarfare is an underappreciated risk that poses a credible threat of catastrophic harm. 

\subsubsection{Risk dimensions}
\begin{itemize}
    \item \textbf{Intent}: Intentional
    \item \textbf{Competency}: Competent
    \item \textbf{Entity}: Variable
    \item \textbf{Polarity}: Single-agent
    \item \textbf{Linearity}: Linear
    \item \textbf{Reach}: Internalized
    \item \textbf{Order}: First-order
\end{itemize}

Similar to CBRN risk, cyber offense represents a broad class of risk that stems from misuse of capable models. However, in contrast to CBRN risks, cyberattacks can take place entirely in the digital domain. In theory, it can be conducted completely by AIs (or AI agents) without any human involvement. The pathway to harm is also less direct, as the resultant harm depends on the target of the attack.

\subsubsection{Risk pathway model}
\begin{itemize}
    \item \textbf{Hazard}: Model with cyber offensive capabilities
    \item \textbf{Event}: A cyber attack is carried out
    \item \textbf{Consequence}: Compromise to critical infrastructure, disruption of supply chain, increased international conflicts etc.
\end{itemize}

Similar to CBRN, here we also designate a model with cyber offensive capabilities as the hazard, as these capabilities are dual-use and may or may not lead to harm. For example, a model with such capabilities can be used for either strengthening cyber defences or conducting cyber attacks. When in the hands of a malicious actor, such models can be used on cyber attacks with potentially catastrophic consequences.

\begin{figure}[htbp]
    \centering
    \includegraphics[width=0.8\textwidth]{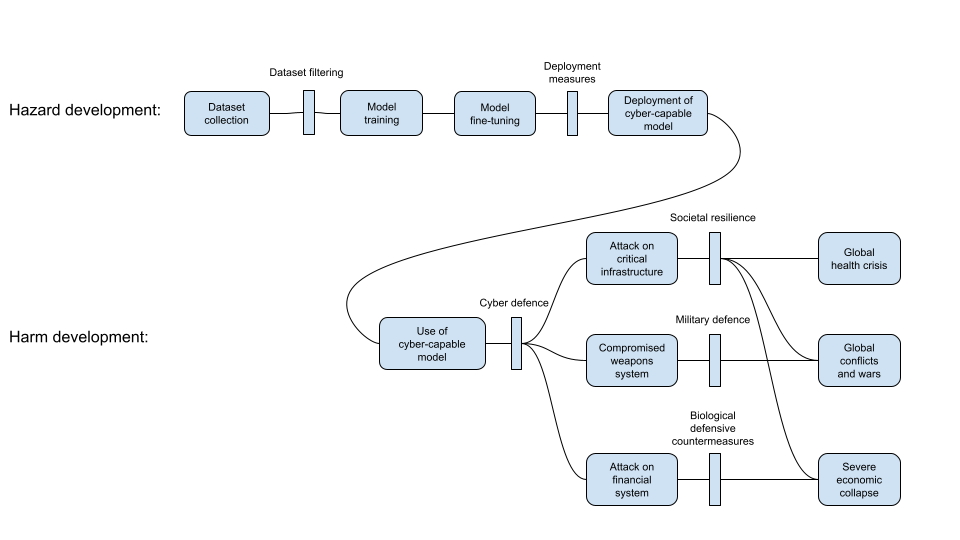}
    \caption{Risk pathway model for a cyber risk scenario}
    \label{fig:cyber_risk}
\end{figure}

In Figure \ref{fig:cyber_risk}, the hazard, a cyber-capable model, is developed and deployed. A malicious actor can then use it to conduct cyber attacks in different areas that could potentially lead to catastrophic consequences. Risk management measures can be placed upstream to reduce the capabilities of these models, or downstream to boost cyber defenses, or even further downstream to limit the damages caused by a cyber attack.

\subsubsection{Analogous historical precedence}
Stuxnet, a worm designed to attack industrial control systems, is considered as the first cyber warfare weapon ever \cite{5772960}, \cite{CERTIST2010Stuxnet}. The Stuxnet malware reportedly caused the damage and subsequent decommissioning of 1000 centrifuges at the Natanz Enrichment Plant, potentially setting back Iran’s progress in its nuclear program \cite{Albright2010Stuxnet}. 

Given that the Stuxnet attack happened in 2010, modern AIs were likely not involved. Nevertheless, it is believed that AIs will increase the volume and heighten the impact of cyber attacks in the near term \cite{NCSC2025AIThreat}.

\subsubsection{Other similar risks}
The development and use of lethal autonomous weapons (LAWs) enables misuse by either humans or, without human involvement, by AIs themselves, with the potential to cause mass casualties \cite{unodaccwbackground}, \cite{humblefutureconflict2024}. 

\subsection{Sudden loss of control}
Sudden loss of control, also known as an AI takeover \cite{balestakeover2024}, is a scenario where an AI rapidly achieves superintelligence through “fast takeoff” or recursive self-improvement. This poses an existential risk \cite{bostromsuperintelligence2014}, \cite{brundagereview2015}. This risk is primarily based on two key ideas: the orthogonality thesis \cite{bostrominstrumental2012}, \cite{armstrongorthogonality2013} and the instrumental convergence thesis \cite{omohundroaidrives2008}, \cite{Shulman2010BasicAIDrives}. Together, these theories argue that a superintelligent AI, regardless of its original goals, would develop power-seeking tendencies as a means to achieve those goals. However, arguments for this scenario typically do not spell out the concrete physical pathways an existential catastrophe would be realized. Instead, they argue that it is the default outcome given the eventual creation of a superintelligence based on a set of reasonable assumptions.

\subsubsection{Risk dimensions}
\begin{itemize}
    \item \textbf{Intent}: Variable
    \item \textbf{Competency}: Competent
    \item \textbf{Entity}: AI
    \item \textbf{Polarity}: Single-agent
    \item \textbf{Linearity}: Linear
    \item \textbf{Reach}: Internalized
    \item \textbf{Order}: First-order
\end{itemize}

The key characteristic of this risk is that a single AI agent competently takes actions that lead to a catastrophic outcome. It does not require the AI to be intentional in its actions, only competent enough to make and execute plans that ultimately result in a catastrophe.

\subsubsection{Risk pathway model}
\begin{itemize}
    \item \textbf{Hazard}: AI with superintelligent capabilities
    \item \textbf{Event}: AI creates bioweapons and build drones
    \item \textbf{Consequence}: Human extinction
\end{itemize}

Here, we designate a superintelligent AI as the hazard. A sufficiently misaligned and competent AI can then carry out plans that lead to human extinction. 

\begin{figure}[htbp]
    \centering
    \includegraphics[width=0.8\textwidth]{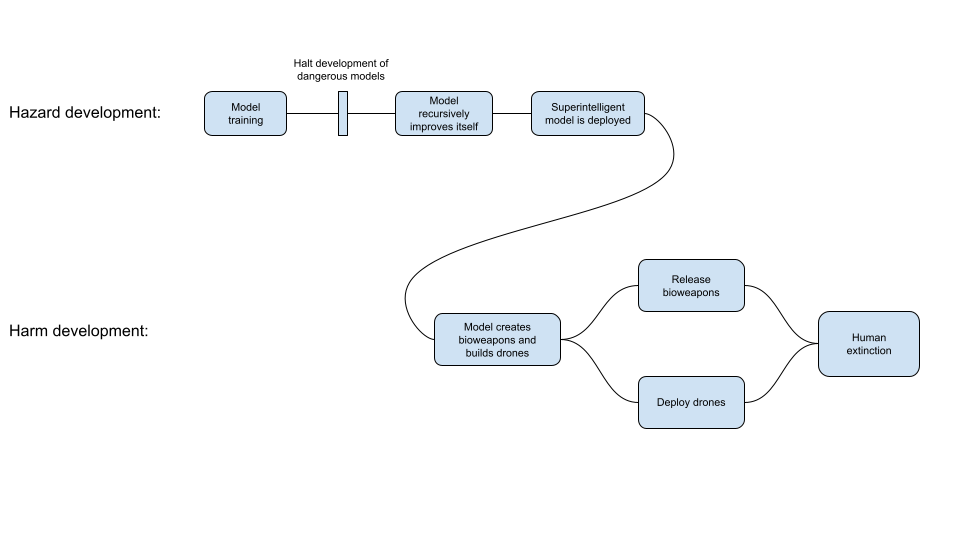}
    \caption{Risk pathway model for a sudden loss of control scenario}
    \label{fig:sudden_loss_of_control}
\end{figure}

We explore a fictitious scenario by \cite{aifuturesmas2025} in Figure \ref{fig:sudden_loss_of_control} where the hazard, a superintelligent AI, is developed and deployed. The model then takes actions in the physical world that leads to catastrophic harm. Risk management measures are primarily put upstream, where model development should be halted appropriately, for example when models are approaching dangerous levels of automated research and development (R\&D).

As this is a work of fiction, some may argue that the exact pathway to harm appears unrealistic. Nevertheless, the focus of this scenario is typically either on preventing models from being superintelligent, or ensuring that models are aligned to human preferences sufficiently that bad outcomes do not get realized if an AI becomes superintelligent. Hence, in the discourse of sudden loss of control risks, there tends to be less focus on generating concrete scenarios and creating risk models.

\subsubsection{Analogous historical precedence}
On 11th September 1973, the democratic socialist president of Chile Salvador Allende and his Popular Unity coalition government was overthrown in a coup d'état by the Chilean military, ending a 46-year history of democratic rule in Chile \cite{DOS_nd_Allende}. Despite Salvador Allende’s Popular Unity party having increased their congressional election votes to 44 percent in March 1973 (up from 36 percent in 1970) merely six months before the coup, there was little he could do to prevent the military from defecting \cite{Bigelow2023ChileTimeline}. This intentional and covertly coordinated subversion was followed by 17 years of military dictatorship which resulted in violent repressions on a massive scale, with tens of thousands reportedly tortured and thousands reportedly killed \cite{CJA_nd_Chile}. 

This historical event illustrates the possibility of an established system losing control in a relatively short timeframe to a powerful and intentional entity with misaligned objectives, a situation that is both difficult to prevent and difficult to reverse. While the evident difference between the 1973 Chilean coup and a sudden AI takeover is that the coup was entirely conducted by humans, it is conceivable for a similar event to be orchestrated by a sufficiently autonomous and misaligned AI agent through remote coordination in the future. These scenarios can be further exacerbated if, similar to a covert human conspiracy, there are minimal detectable signs of an impending AI takeover, thereby hampering preventive action; and when the adversary rapidly gains power over a short timeframe, as per the AI fast takeoff scenario, giving little time for any meaningful preparation. It has been argued that a sufficiently powerful AI with scheming capabilities would be able to pursue unintended objectives that are difficult to detect or intervene, leading to a loss of control scenario \cite{balesni2024evaluationsbasedsafetycasesai}.

\subsection{Gradual loss of control}
Gradual or accumulative loss of control risks can be described as risks resulting from the accumulation of less severe disruptions that gradually weakens systemic resilience until a critical event triggers a catastrophe \cite{kasirzadeh2025typesaiexistentialrisk}, \cite{krook2025autonomybreakshiddenexistential}.

\subsubsection{Risk dimensions}
\begin{itemize}
    \item \textbf{Intent}: Unintentional
    \item \textbf{Competency}: Variable
    \item \textbf{Entity}: Variable
    \item \textbf{Polarity}: Multi-agent
    \item \textbf{Linearity}: Non-linear
    \item \textbf{Reach}: Internalized
    \item \textbf{Order}: Variable
\end{itemize}

The key characteristics of this risk is that it is not caused by a single agent leading to a single defining event, instead, it is primarily about its multi-agentic and non-linear nature, where the deep integration of AIs into society leads to structural and systemic weakness. 

\subsubsection{Risk pathway model}
\begin{itemize}
    \item \textbf{Hazard}: AIs with general capabilities
    \item \textbf{Event}: AI displaces human labour
    \item \textbf{Consequence}: Humans lose autonomy
\end{itemize}

A hazard like AIs with general capabilities may be viewed positively due to its potential societal benefits. However, in this risk pathway, this hazard could lead to the event of AI displacing human labor, which can result in humans losing autonomy. 

The flowchart below is based on a scenario of gradual disempowerment, describing the transition to an AI-dominated economy, as depicted by \cite{kulveit2025gradualdisempowermentsystemicexistential}. 

\begin{figure}[htbp]
    \centering
    \includegraphics[width=0.8\textwidth]{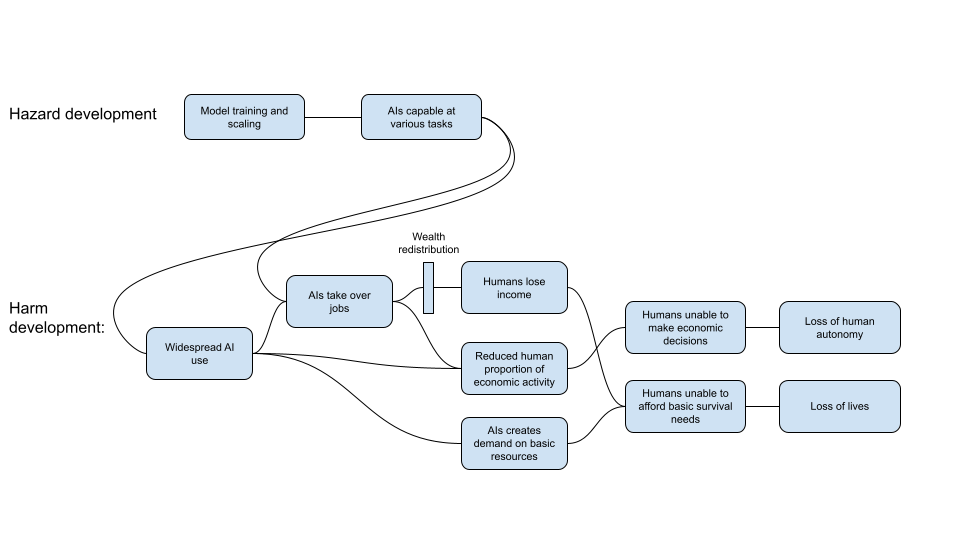}
    \caption{Risk pathway model for a gradual loss of control scenario}
    \label{fig:gradual_loss_of_control}
\end{figure}

In Figure \ref{fig:gradual_loss_of_control}, gradual loss of control happens when AI capability leads to its widespread use, consequently displacing humans from economically viable jobs and leaving humans unable to afford basic survival needs. Assuming the hazard is AIs capable at various tasks, risk management is difficult to be performed upstream, as this dual-use hazard is largely desirable. Much of the risk management would then need to be performed downstream, both in terms of managing the integration of AI into society, as well as ensuring that the socioeconomic needs of people are fulfilled.

\subsubsection{Analogous historical precedence}
On 6th May 2010, in an incident later known as the 2010 Flash Crash, leading U.S. stock indices abruptly fell and rebounded in less than half an hour, in the process erasing almost \$1 trillion in market value. An investigation by the Security Exchange Commission found that a single order of large amounts of E-mini S\&P contracts and subsequent selling orders by high-frequency algorithms triggered the drastic decline of market value \cite{SEC_CFTC2010MarketEvents}, \cite{Kirilenko2017FlashCrash}. This event demonstrated the problem of algorithmic collision, where an increasing deployment of algorithms interacting with each other can lead to unforeseen and catastrophic consequences \cite{chiodo2025problemalgorithmiccollisionsmitigating}. 

While the crash did not result in a loss of human autonomy or lives, it serves as a historical precedent for gradual loss of control, where it showed that risks do not require a sudden "fast takeoff" in AI capabilities. Instead, a gradual diffusion and deep entrenchment of AI into a system can lead to a sudden and catastrophic event. In the case of the 2010 Flash Crash, although the market collapsed in a matter of minutes, the transition to automated high-frequency trading took years.

\subsection{Environmental risk}
AI models are often trained using large amounts of computation. This process is very energy intensive, potentially leading to significant greenhouse emissions depending on the energy sources \cite{ebert2025aiclimateregulationdata}. Experts believe drastically increasing carbon emissions could accelerate climate change, which may constitute a catastrophic risk \cite{ipccsynthesis2023}.

\subsubsection{Risk dimensions}
\begin{itemize}
    \item \textbf{Intent}: Unintentional
    \item \textbf{Competency}: Variable
    \item \textbf{Entity}: Variable
    \item \textbf{Polarity}: Variable
    \item \textbf{Linearity}: Linear
    \item \textbf{Reach}: Externalized
    \item \textbf{Order}: First-order
\end{itemize}

The key characteristic of environmental risks resulting from AI is that it is an externality, where those who suffer from the outcome include third parties who are not directly part of the value chain.

\subsubsection{Risk pathway model}
\begin{itemize}
    \item \textbf{Hazard}: Energy-intensive data centers
    \item \textbf{Event}: Increasing usage of carbon intensive energy sources
    \item \textbf{Consequence}: Environmental harm
\end{itemize}

In contrast to the previous risks, this hazard is not tied to AI model capabilities. Here, the hazard is energy-intensive data centers, which can lead to increased carbon emissions if they consume carbon-intensive energy sources. Because this risk is realized cumulatively over time, there is no single event that triggers the harm; it is a continuous process.

\begin{figure}[htbp]
    \centering
    \includegraphics[width=0.8\textwidth]{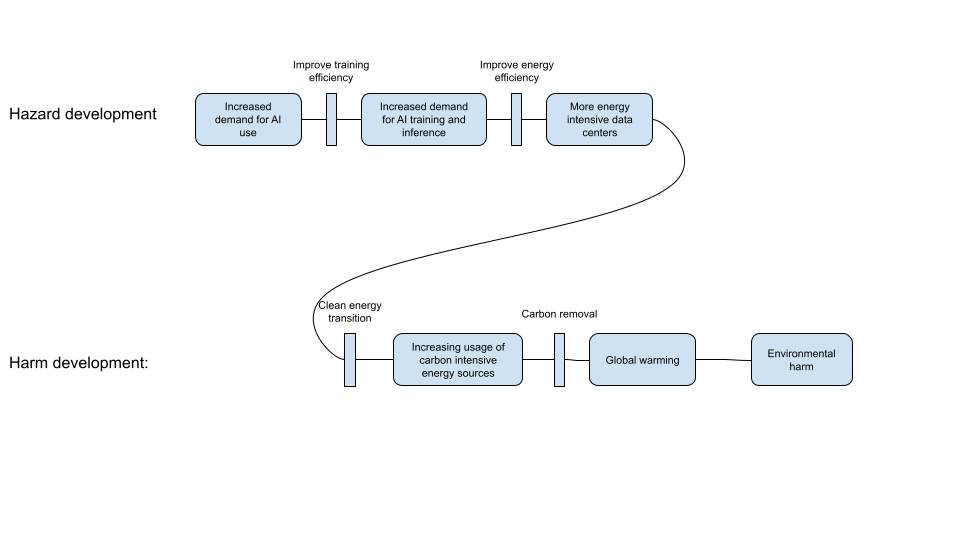}
    \caption{Risk pathway model for an environmental risk scenario}
    \label{fig:environmental_risk}
\end{figure}

In Figure \ref{fig:environmental_risk}, an increased demand for AI leads to the hazard of more energy-intensive data centers. When these centers rely on carbon-intensive energy sources, the result can be environmental harm. Upstream risk management measures include making model training and data centers more energy-efficient, whereas downstream measures focus on ensuring a supply of clean energy and implementing carbon removal strategies.

\subsubsection{Analogous historical precedence}
In 1974, \cite{Molina1974} proposed that stratospheric ozone might be destroyed by industrially produced substances including chlorofluorocarbons (CFC) which are commonly used in refrigerators and air conditioners. This ozone depletion is believed to have led to an increase in global skin cancer prevalence through overexposure to the sun, posing a significant world-wide health burden \cite{Norval2011}. To manage this externality, the Montreal Protocol, a global agreement to phase out chemicals that led to the ozone depletion, was eventually signed in 1987 and entered into force in 1989 \cite{UNEP_nd_MontrealProtocol}.

Prior to the Montreal Protocol, the use of CFC-based household appliances such as refrigerators and air conditioners inadvertently led to the destruction of the environment, adversely impacting people who were not directly part of the value chain of these appliances. Similarly, there is a risk of externalities from AI posing environmental harms. It is reported that the combined footprint of the leading 200 digital companies represents 0.8\% of all global energy-related emissions, with a significant proportion of it coming from data centres that power AI \cite{ITU2025GreeningDigital}.

\subsection{Geopolitical risk}
As AI is increasingly seen as a powerful technology, countries are racing to develop it ahead of their geopolitical rivals, a competition that could lead to geopolitical tensions \cite{alqutbahrivalries2025}, \cite{barnett2025aigovernanceavoidextinction}.

\subsubsection{Risk dimensions}
\begin{itemize}
    \item \textbf{Intent}: Unintentional
    \item \textbf{Competency}: Variable
    \item \textbf{Entity}: Variable
    \item \textbf{Polarity}: Variable
    \item \textbf{Linearity}: Non-linear
    \item \textbf{Reach}: Externalized
    \item \textbf{Order}: Second-order
\end{itemize}

The emphasis of this risk is on harms that result from second-order effects, where geopolitical instabilities result from the race to develop AI, rather than on the direct consequences of the deployment or use of AI itself.

\subsubsection{Risk pathway model}
\begin{itemize}
    \item \textbf{Hazard}: Destabilized geopolitical environment
    \item \textbf{Event}: Could be triggered by any event
    \item \textbf{Consequence}: International conflict and wars
\end{itemize}

For this risk, the designation of a hazard and event is less straightforward, primarily because it is a second-order effect. Unlike other hazards that can be neutral, a destabilized geopolitical environment is inherently undesirable. Furthermore, the mechanism for hazard release is difficult to predict, as minor unexpected triggers can rapidly escalate into larger events.

\begin{figure}[htbp]
    \centering
    \includegraphics[width=0.8\textwidth]{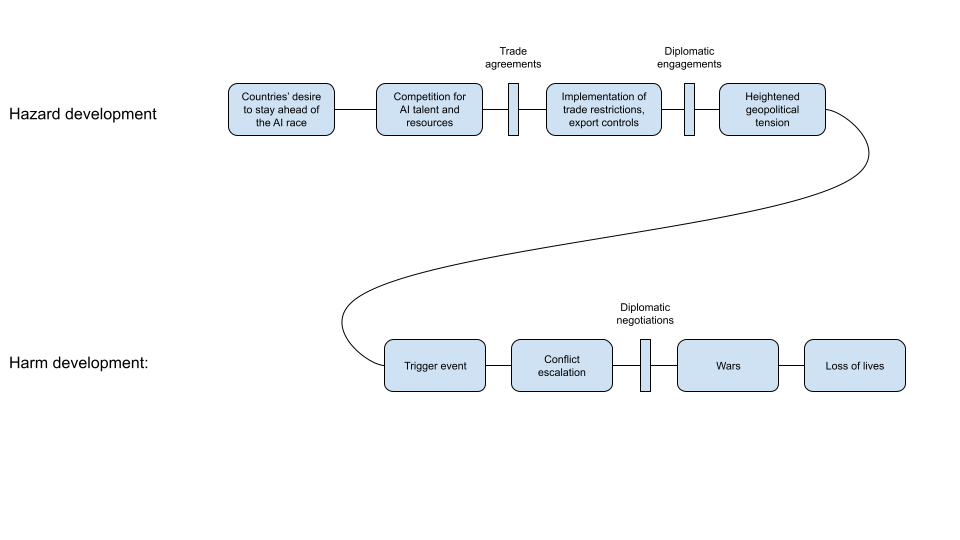}
    \caption{Risk pathway model for a geopolitical risk scenario}
    \label{fig:geopolitical_risk}
\end{figure}

In Figure \ref{fig:geopolitical_risk}, an AI race leads to heightened geopolitical tension which results in a war with catastrophic consequences. Risk management measures here are thus less relevant to AI development and deployment itself, but rather more focused on managing geopolitical relationships. 

\subsubsection{Analogous historical precedence}
Unlike many other wars where states fought over land and resources, the Cold War was primarily an ideological confrontation, where both the U.S. and the Soviet Union sought to establish global supremacy of their desired political and economic models. Though it did not result in direct military engagement between the two major powers, this conflict frequently led to widespread proxy wars across various regions such as Vietnam and Afghanistan \cite{Venkatraj2019RisingConflicts}. While the causes of these proxy wars were often rooted in complex local and regional dynamics, their scale and intensity were significantly exacerbated as a second-order effect of the overarching ideological conflict between the two superpowers.

Instead of an ideological race, some countries are now focusing on winning the AI race instead. For example, the U.S. AI and Crypto Czar David Sacks was quoted as saying “to remain the leading economic and military power, the United States must win the AI race” \cite{WhiteHouse2025AIAction}. This strategic rivalry, particularly between the U.S. and China, is discussed by \cite{Esposito2025AIGeopolitics} as a new “digital Cold War”. Experts have also warned that certain AI-related attacks could tip the geopolitical sphere that result in conditions similar to those that preceded the First and Second World Wars \cite{Campbell2022AIColdWar}.

\subsubsection{Other similar risks}
It is argued that the widespread use of AIs can have potentially catastrophic epistemic side effects, where AIs system may make changes to humans’ (or other agents’) knowledge or beliefs because it was not told not to do so \cite{Klassen2023EpistemicSideEffects}. Similarly, \cite{wright2025cognitivecastesartificialintelligence} argues that AIs could lead to “structural violence” where due to unequal distribution of epistemic vulnerability, there will be a divide between epistemic agency and epistemic automation, where the vast majority of people who lack logic and scrutiny will cease to reason. 

\section{Summary}
Table \ref{tab:summary} summarizes the risk dimensions for each of the risks discussed. These risks are not meant to be comprehensive nor exhaustive, but they serve to illustrate a broad class of risk with at least one of the dimensions being distinct from the other risks. 

\begin{table}[h!]
    \centering
    \caption{Summary of risk dimensions for the risks discussed}
    \label{tab:summary}
    \begin{tabular}{|p{2cm}|l|l|l|p{1.5cm}|l|l|p{1.3cm}|}
        \hline
        \textbf{Risks} & \textbf{Intent} & \textbf{Competency} & \textbf{Entity} & \textbf{Polarity} & \textbf{Linearity} & \textbf{Reach} & \textbf{Order} \\
        \hline
        CBRN & Intentional & Competent & Humans & Single-agent & Linear & Internalized & First-order \\
        \hline
        Cyber offense & Intentional & Competent & Variable & Single-agent & Linear & Internalized & First-order \\
        \hline
        Sudden loss of control & Variable & Competent & AI & Single-agent & Linear & Internalized & First-order \\
        \hline
        Gradual loss of control & Unintentional & Competent & Variable & Multi-agent & Non-linear & Internalized & First-order \\
        \hline
        Environmental risks & Unintentional & Variable & Variable & Variable & Linear & Externalized & First-order \\
        \hline
        Geopolitical risks & Unintentional & Variable & Variable & Variable & Non-linear & Externalized & Second-order \\
        \hline
    \end{tabular}
\end{table}

While these risks are analyzed independently to and characterized according to their risk dimensions, real-world risks are far more complex and interconnected. For example, an AI race that leads to geopolitical risks may result in an authoritarian regime carrying out an AI-enabled cyberattack on an autonomous weapons system. In this example, multiple risks with different risk dimensions may play out simultaneously, resulting in a "polycrisis" \cite{Lawrence2024PolycrisisRoadmap}. In addition, rapid advances in domains outside of AI such as biotech, robotics, quantum computing, and energy may also amplify risks \cite{suleymancomingwave2023}. In response, fields of studies at the intersection of different risks have emerged, such as biocybersecurity (or cyberbiosecurity), which attempts to address both cyber and biological risks \cite{potter2020biocybersecurityconvergingthreat}.

\section{Limitations}
While this paper proposes an approach to analyzing catastrophic AI risks through dimensional characterization and risk pathway modeling, several limitations should be acknowledged:

\begin{itemize}
    \item \textbf{Non-exhaustiveness of risks:} The six risks explored are not intended to be comprehensive, as they were selected because they are commonly discussed, where each risk has a distinct combination of risk dimensions. While we have included other examples of risks with similar dimensions where applicable, they remain non-exhaustive.
    \item \textbf{Subjectivity of risk dimensions:} The seven risk dimensions discussed alongside their attributes are just several ways AI risks can be characterized, such that risk management measures broadly associated with those attributes can be identified. There may be other useful dimensions that are not captured in this analysis.
    \item \textbf{Incompleteness of risk pathways:} The risk pathways shown only represent a particular scenario for a given risk, and does not intend to cover a significant distribution of how risks are realized. There may be other plausible scenarios that have not been covered by this analysis. 
    \item \textbf{Lack of quantitative modeling:} This work focuses on qualitative causal mapping rather than quantitative risk estimation. As such, it does not attempt to assign probabilities nor severity to these risks.
\end{itemize}

\section{Conclusion}
Our work has provided a framework for characterizing AI risks in terms of their risk dimensions, where we explored six commonly discussed catastrophic risks along seven dimensions. For each attribute of these dimensions, the relevant risk management measures are listed, where they can be applied to risks associated with those dimensions.

We also conducted simple risk pathway modeling to map out the causal path to harm for each scenario. While we sometimes identified model capabilities as the hazard, many of the risks explored lie outside the control of model developers. Though the least cost avoidance principle correctly suggests focusing on the upstream parts of the AI value chain, we believe that an upstream focus is a necessary but insufficient condition for good risk management. It is crucial to conduct risk management at all levels of the value chain, which includes:
\begin{itemize}
    \item The model level, e.g. during data collection, training, post-training, or deployment \cite{shevlane2023modelevaluationextremerisks}.
    \item The system or application level i.e. the specific areas it is being used \cite{goh2025measuringmattersframeworkevaluating}
    \item Broader societal and political level \cite{Maslej2025AIIndex}, \cite{bernardi2025societaladaptationadvancedai}
\end{itemize}

Ultimately, a comprehensive strategy for managing catastrophic AI risks requires both a multi-dimensional understanding and concrete causal mapping. These can then facilitate robust risk management implemented across all levels of the AI value chain.

Future research can focus on four key areas. First, risks can be identified by conducting open-ended exploration across various risk dimensions. Second, more comprehensive risk pathway models can be developed, incorporating a wider array of scenarios and detailed mitigation strategies. Third, risks can be quantified by estimating the probability and severity based on its pathways and the effectiveness of proposed mitigations. Lastly, further investigation into risk mitigation measures beyond the model and application level, at the application and societal level, would be beneficial for managing these risks holistically.

\section{Acknowledgements}
This work was funded by the Open Philanthropy AI Governance and Policy grant, and supported by LISA (London Initiative for Safe AI) through subsidized access to its research coworking space. We also thank Francesca Gomez, Ben Bucknall, Marta Ziosi, and Peter Slattery for helpful discussion and feedback.

\clearpage
\printbibliography
\end{document}